\documentclass{aa}  

\usepackage{graphicx}
\usepackage{txfonts}
\usepackage{amsmath}
\usepackage[skip=0.333\baselineskip]{caption}
\usepackage{subcaption}
\usepackage{xcolor} % for color
\usepackage{ulem} % for sout

\usepackage{natbib}

\begin{document}

   \title{SWEET-Cat: A view on the planetary mass-radius relation\thanks{A copy of the SWEET-Cat catalog used for this work is available in electronic form at the CDS via anonymous ftp to cdsarc.u-strasbg.fr (130.79.128.5) or via http://cdsweb.u-strasbg.fr/cgi-bin/qcat?J/A+A/.}}

%   \subtitle{The density of different populations of planets}

   \author{S. G. Sousa\inst{1}                  %leader of paper
          \and V. Adibekyan\inst{1}             % discussion, many comments   
          \and E. Delgado-Mena\inst{1}
          \and N. C. Santos\inst{1,}\inst{2}
          \and B. Rojas-Ayala\inst{3}           
          \and S. C. C. Barros\inst{1}          %
          %\and M. Deleuil\inst{4} 
          \and O. D. S. Demangeon\inst{1,}\inst{2}
          \and S. Hoyer\inst{4}                 %PI of SOPHIE program
          \and G. Israelian\inst{5}
          \and A. Mortier\inst{6}            
          \and B. M. T. B. Soares\inst{1,}\inst{2}  %
          \and M. Tsantaki\inst{7}                
          }

          \institute{Instituto de Astrof\'isica e Ci\^encias do Espa\c{c}o, Universidade do Porto, CAUP, Rua das Estrelas, 4150-762 Porto, Portugal
          \and Departamento de F\'isica e Astronomia, Faculdade de Ci\^encias, Universidade do Porto, Rua do Campo Alegre, 4169-007 Porto, Portugal
          \and {Instituto de Alta Investigaci\'on, Universidad de Tarapac\'a, Casilla 7D, Arica, Chile}
%          \and Department of Astronomy, University of Geneva, 51 chemin Pegasi, 1290 Sauverny
          \and Aix Marseille Univ, CNRS, CNES, LAM, Marseille, France
          \and Instituto de Astrof\'isica de Canarias, 38200 La Laguna, Tenerife, Spain
%          \and European Southern Observatory, Alonso de Cordova 3107, Vitacura, Santiago, Chile
%          \and Centre for Exoplanet Science, SUPA, School of Physics and Astronomy, University of St Andrews, St Andrews KY16 9SS, UK
%          \and Dipartimento di Fisica e Astronomia "Galileo Galilei", Universit\'a di Padova, Vicolo dell'Osservatorio 3, Padova IT-35122, Italy
          \and School of Physics \& Astronomy, University of Birmingham, Edgbaston, Birmingham, B15 2TT, UK
          \and INAF -- Osservatorio Astrofisico di Arcetri, Largo Enrico Fermi 5, 50125 Firenze, Italy
          }

   \date{Received September 15, 1996; accepted March 16, 1997}

% \abstract{}{}{}{}{} 
% 5 {} token are mandatory
 
  \abstract
  % context heading (optional)
  % {} leave it empty if necessary  
   {}
  % aims heading (mandatory)
   {SWEET-Cat (Stars With ExoplanETs Catalogue) was originally introduced in 2013, and since then, the number of confirmed exoplanets has increased significantly. A crucial step for a comprehensive understanding of these new worlds is the precise and homogeneous characterization of their host stars. %This paper is one more step for the increase of the number of planet-host stars with homogeneous stellar parameters.
   } 
  % methods heading (mandatory)
   {We used a large number of high-resolution spectra to continue the addition of new stellar parameters for planet-host stars in SWEET-Cat following the new detection of exoplanets listed both at the Extrasolar Planets Encyclopedia and at the NASA exoplanet archive. We obtained high-resolution spectra for a significant number of these planet-host stars, either observed by our team or collected through public archives. For FGK stars, the spectroscopic stellar parameters were derived for the spectra following the same homogeneous process using ARES+MOOG as for the previous SWEET-Cat releases. The stellar properties are combined with the planet properties to study possible correlations that could shed more light into the star-planet connection studies.
   }
  % results heading (mandatory)
   {We increase the number of stars with homogeneous parameters by 232 ($\sim$ 25\% - from 959 to 1191). We then focus on the exoplanets with both mass and radius determined to review the mass-radius relation where we find consistent results with the ones previously reported in the literature. For the massive planets we also revisit the radius anomaly where we confirm a metallicity correlation for the radius anomaly already hinted in previous results.
   }
  % conclusions heading (optional), leave it empty if necessary 
   {}


   \maketitle
%
%-------------------------------------------------------------------
%Vardan: I would more focus on the importance of homogeneity.. Give some examples how the results can be improved if the sample has homogeneous parameters etc.
%Annelies: Add paragraph about importance of homogeneous stellar parameters, with focus on what follows in the discussion.
%
\section{Introduction}

Planetary systems studies have grown immensely in the last couple of decades following the first detections of exoplanets orbiting solar-type stars \citep[e.g.,][]{Mayor-1995, Marcy-1996} and the thousands of exoplanets detected since. More interesting than the fantastic increase rate of discoveries is the surprising vast diversity of exoplanets present in the Galaxy, unraveled mostly by dedicated search surveys using the radial-velocity (RV) and transit techniques \citep[e.g.][]{Ehrenreich-2020, Lillo-Box-2020, Armstrong-2020}. These discoveries provide crucial constraints for the exhaustive understanding of planet formation and evolution, based initially on our knowledge of the solar system alone.

Because the main techniques to detect and characterize these exoplanets rely on the observation of their host stars, it is crucial to derive precise, accurate, and homogeneous stellar parameters. The homogeneous determination of such parameters is even more important if we want to perform statistical studies on the different populations of exoplanets that have been discovered in the last years \citep[e.g.][]{Adibekyan-2024, Buchhave-2018, Santos-2017, Goda-2019}.

The SWEET-Cat database (“Stars With ExoplanETs Catalogue” \footnote{sweetcat.iastro.pt}) compiles an updated list of all stars known to host planets. First presented in \citet[][]{Santos-2013}, SWEET-Cat started with a list of only $\sim$700 planet-host stars, with about 65\% of these having homogeneous spectroscopic parameters derived by our team using a well defined spectral analysis procedure. The number of planet-host stars has grown to more than 4000 stars (hosting approximately 5000 exoplanets in total) during the past decade given the very productive results of many well known ground and space based projects such as HARPS, ESPRESSO, \textit{Kepler}, and TESS \citep[][respectively]{Mayor-2011, Pepe-2021, Borucki-2010, Ricker-2014}. The high rate of planet-host discoveries is hard to follow with the derivation of homogeneous spectroscopic parameters, specially for the many faint stars coming from \textit{Kepler} transit detections for which it is much more difficult to obtain high quality spectra. Because of this, in the last years the percentage of planet-host stars in SWEET-Cat with homogeneous parameters have decreased to about only 20\%.

Here we present a significant update to SWEET-Cat. This work consists of the addition of 232 new stellar atmospheric parameters for planet-hosts, in addition to the 31 recent planet-host stars which we also included in SWEET-Cat since its last update in 2021. These came from different planet detection and characterization works in which our team has contributed for the stellar characterization. For all these new planet-host stars we maintain the link to the NASA exoplanet archive database\footnote{http://exoplanetarchive.ipac.caltech.edu}\citep[][]{Akeson-2013}, as well as the Extrasolar Planets Encyclopedia\footnote{http://exoplanet.eu/}\citep[][]{Schneider-2011} to allow an easy correlation between stellar and planet properties for the community. 

The following sections describe the work done for the update of SWEET-Cat as well as a short review of the mass-radius of exoplanets as an example for the use of SWEET-Cat data. Section 2 describes the spectroscopic data compilation for the catalog. Section 3 recalls how the spectroscopic parameters are homogeneously derived for SWEET-Cat. Section 4 is focused on the planetary mass and radius relations review. In Section 5 we revisit the radius anomaly observed for giant massive planets and its observational link with metallicity. In Section 6 we conclude with a summary of this work.

%--------------------------------------------------------------------
\section{Spectroscopic Data}

\subsection{Selection of host stars for SWEET-Cat}

We used the same procedure as before to add new planet-host stars in SWEET-Cat. We start by searching all confirmed planets in the Extrasolar Planets Encyclopedia and select the detected by ``Radial Velocity'', ``Primary Transit'', and ``Astrometry'' methods, only. We then perform a cross-match of these with the stars in SWEET-Cat and include the missing hosts while at the same time we remove some of the stars for which the previous planet detection is controversial at present. The same is performed with the NASA exoplanet archive to add the confirmed host stars missing in the Extrasolar Planets Encyclopedia. We also keep the few confirmed planet-hosts that are not in common in between the exoplanets databases. The source of the planet's database is also updated in SWEET-Cat with this information.

\subsection{Spectra compilation}

The compilation of the spectra follows the same exact procedure as described in section 2.2 from \citet[][]{Sousa-2021}. Here we briefly summarize the procedure. We use the Spectral Data Product Form of the ESO Archive\footnote{http://archive.eso.org/wdb/wdb/adp/phase3\_main/form} to automatically find and download the data we use the \textit{astroquery.eso}\footnote{http://astroquery.readthedocs.io} sub-module.
This search includes private data from our recent UVES programs for SWEET-Cat (106.20ZM.001, 109.22VY.001, 110.23TD.001, and 111.24HZ.001) as well as high quality publicly available data of the host stars with declinations less than $\sim +30^{\circ}$.

For each FGK star (with effective temperature above 4500K and below 7000K \footnote{For M stars a different spectroscopic analysis is used and is addressed in Antoniadis-Karnavas et al. 2024, accepted A\&A}) listed in SWEET-Cat we searched for individual exposures of high resolution spectra and combined these individual exposures for each instrument (and different configuration, in the case of UVES). The combination of the spectra is done by adding all the individual spectra after the respective radial velocity shift which is estimated during this process using one of the exposures as a reference spectra.

We also searched spectra in other public archives including data from the SOPHIE\footnote{http://atlas.obs-hp.fr/sophie/} and HARPS-N\footnote{http://archives.ia2.inaf.it/tng/} spectrographs\footnote{Other instruments and public archives are planned to be included in future works depending on the automatization to retrieve and treat the data from them.}. We combined these data in the same way as for the ESO spectral data. Here we would also like to emphasize a specific program (20B.PNP.Hoyer) which was run on SOPHIE to collect planet-host spectra from the northern hemisphere. 

For this work we compiled a total of 333 combined spectra for 251 different FGK stars. The typical S/N is around 150-200 where only 18\% of these spectra is below a S/N $\sim$ 100.

\section{Stellar Parameters}

\begin{figure*}
  \centering
  \includegraphics[width=0.95\linewidth]{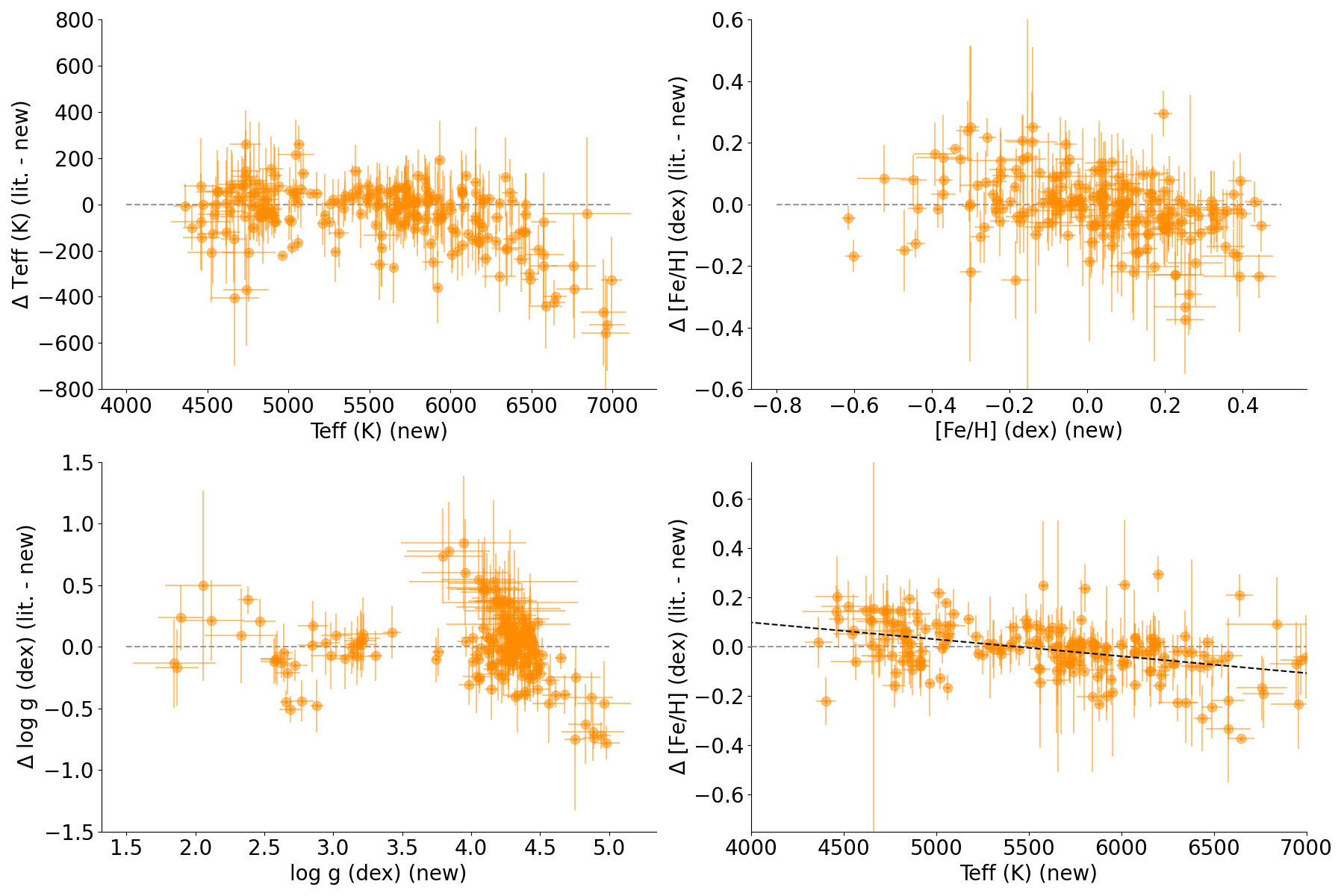}
  \caption{The new SWEET-Cat homogeneous parameters ($T_\mathrm{eff}$, $\log g$, and [Fe/H]) compared to the literature values previously listed in SWEET-Cat. The right bottom panel shows the difference of [Fe/H] vs. the derived effective temperature. Here the dashed black line represents a linear fit.}
  \label{fig_param}
\end{figure*}

\subsection{Spectroscopic parameters}

We carried out the spectroscopic analysis with ”ARES+MOOG“ for the new spectroscopic data. The spectral analysis relies on the excitation and ionization balance of iron abundance. The ARES code\footnote{The latest version of ARES can be found on github.com/sousasag/ARES} \citep[][]{Sousa-2007, Sousa-2015b} automatically measured the equivalent widths of the absorption lines. The radiative transfer code MOOG  \citep[version of Nov. 2019;][]{Sneden-1973} was used for the element abundances assuming local thermodynamic equilibrium (LTE) and using a grid of Kurucz ATLAS9 plane-parallel model atmospheres \citep[][]{Kurucz-1993}. We applied this method in our previous spectroscopic studies of planet-hosts \citep[e.g.,][]{Sousa-2008, Sousa-2011, Mortier-2013b, Sousa-2015a, Andreasen-2017, Sousa-2018, Sousa-2021}. We use the same line list introduced in \citet[][]{Sousa-2008}, except for the stars with effective temperature below 5200K where we used the line list provided in \citet[][]{Tsantaki-2013}. %\footnote{The log gf for these lines were recalibrated for a proper differential analysis with the latest version of MOOG. There were only very small differences in these recalibrated loggf values.}

For a few stars we were able to collect combined spectra from two or more instruments/configurations. We recall that the results derived from spectra from different instruments are generally very consistent within the errors as shown in \citet[][]{Sousa-2018}. We include only one set of converged parameters per star in SWEET-Cat following the same reasoning as in our last update \citep[][]{Sousa-2021} which is mostly based on the errors derived in the spectroscopic analysis. 

In Figure \ref{fig_param} we show the comparison between the parameters derived in this work and the values previously listed in SWEET-Cat which were compiled previously from literature values.

%  delta_Teff:  44.37 +- 132.90 K
%  delta_Logg:   -0.00 +-  0.27 dex
%  delta_[Fe/H]:   0.01 +-  0.11 dex
We derived homogeneous parameters for 232 new stars in SWEET-Cat. For nineteen stars in our starting sample convergence was not reached because either the spectra were of bad quality, or the star has a relatively high rotation rate. Both these issues cause difficulties to obtain good measurements of the lines's equivalent widths.
General consistency between our values and the literature (mainly gathered from the discovery paper) is similar to previous comparisons: for effective temperature and [Fe/H] (top panels of Figure \ref{fig_param}), the mean differences are 44 $\pm$ 133 K and 0.01 $\pm$ 0.11 dex, respectively. There are however significant differences between the parameters for some host stars, in particular for the hotter stars above $\sim$ 6500 K. The offset that we see for these hotter stars is mostly related with the increase of rotational velocity at these temperature which makes it more difficult to derive spectroscopic parameters with different methodologies. For these fourteen stars for which we derived effective temperatures above 6500K all have vsin i values reported in literature between 10-20 km/s. For several of the stars we can find large spread in effective temperature from different methods. One example can be seen in \citet[][]{Raynard-2018} that used three different methods to characterize NGTS-2 for which they derived effective temperatures as high as 6604 $\pm$ 134 K, but in the end they adopt a lower value of $6478^{+94}_{-89}$ K. Other factors such as lower quality of spectra data can also contribute for the offset and spread. The bottom line is that caution should be taken when using different estimations from different methods in particular for these hotter stars.

The metallicity comparison shows also a general consistent result with some outliers. Some of these outliers can even be explained by typos in the exoplanets databases\footnote{As an example, at the time of compilation of these values, for WASP-137 exoplanets.eu listed a metallicity ([Fe/H]) of 0.487 ($\pm$ 0.051) pointing to the work \citet[][]{Anderson-2018} where the value listed is 0.08 $\pm$ 0.09 dex. The 0.487 value was mistakenly sourced from the line below corresponding to the stellar luminosity.}. Moreover when we compare the difference between the derived metallicity and literature values as a function of the effective temperature (bottom right panel of Figure \ref{fig_param}) there is evidence for a small trend ($-6.84\mathrm{e}^{-5} \pm 1.07\mathrm{e}^{-5}$, with a Spearman correlation coefficient of $-0.43$ and a very low p-value of $1.79\mathrm{e}^{-10}$). The observed difference is generally covered by the provided errors as we also see in the top right panel of the same figure, but it is a different way to show the importance of using homogeneous spectral analysis in statistical studies.

%[[Fit Statistics]] for the line of feh differences vs. teff
%    # fitting method   = leastsq
%    # function evals   = 9
%    # data points      = 205
%    # variables        = 2
%    chi-square         = 1.90047493
%    reduced chi-square = 0.00936195
%    Akaike info crit   = -955.585763
%    Bayesian info crit = -948.939743
%[[Variables]]
%    m: -6.8480e-05 +/- 1.0773e-05 (15.73%) (init = 0)
%    b:  0.37217970 +/- 0.06066037 (16.30%) (init = 0)
%[[Correlations]] (unreported correlations are < 0.100)
%    C(m, b) = -0.994
%-6.8479585676684e-05 0.37217969664726525
%spearman statistic, pvalue:
%(-0.42662173449656954, 1.789338500186793e-10)

%SSousa: Confirm that we are using the spectrosocpic logg and not the trigonometric logg in figure 1. Done. We are showing the spectroscopic surface gravity.
In the bottom left panel of Figure \ref{fig_param} the comparison of the spectroscopic surface gravity is generally consistent with a mean difference of 0.00 $\pm$ 0.27 dex, although there are a few outliers with significant differences and a large spread of values is seen for the dwarf stars. This is similar to our previous results and can be corrected at some degree as discussed in \citet[][]{Mortier-2014}.  

We also derive the trigonometric surface gravity following the same procedure as in \citet[][]{Sousa-2021} to obtain more accurate values for all the stars. We matched the new stars with homogeneous parameters listed with the Gaia ID in DR3 using their coordinates and the VizieR catalogues \citep[][]{GAIA-2016, GAIA-2021}. We checked the magnitudes and astrometry data to confirm that we selected the correct star. For each target with GAIA ID, we derived the stellar luminosities directly from the precise data in GAIA DR3, following the equation 8.6 from the GAIA documentation\footnote{https://gea.esac.esa.int/archive/documentation/GDR2/ - in particular chapter 8.3.3. authored by Orlagh Creevey and Christophe Ordenovic} where $M_G$ is the absolute GAIA magnitude, $T_\mathrm{eff}$ is the effective temperature listed in SWEET-Cat, $M_\mathrm{bol\odot} = 4.74$ (as defined by the IAU resolution 2015 B2), and $BC_G(T_\mathrm{eff})$ is a temperature-dependent bolometric correction provided in the GAIA documentation \citep[equation 8.9 and table 8.3][]{Andrae-2018}. Instead of a simple inversion of the GAIA parallax to provide a distance, we preferred to include the geometric distance and an error directly taken from the maximum of the asymmetric uncertainty measures (16th and 84th percentiles) reported in \citet[][]{Bailer-Jones-2021}. 

To estimate the trigonometric surface gravity, we need the stellar mass. As in previous works, we used the stellar mass calibration in \citet[][]{Torres-2010}, and for estimates between 0.7 and 1.3M$_\odot$ we used the recommended equation (1) in \citet[][]{Santos-2013} that corrects for the offset between the calibrated stellar mass and the ones estimated using isochrones. However, the \citet[][]{Torres-2010} calibration requires spectroscopic parameters of the star as input, including the surface gravity. Therefore, we performed an iterative process to converge simultaneously to the best estimates of the stellar mass and the trigonometric surface gravity. The procedure only takes a few iterations until it converges and it is described in detail in section 3.4 in \citet[][]{Sousa-2021}.

\begin{table}[t]
\caption[]{Updated statistics for SWEET-Cat.}
  \begin{center}

    \begin{tabular}{ll}
    \hline
    \hline
    Number     &   Description  \\
    \hline

4106 & stars in SWEET-Cat  \\
1191 & planet-hosts with homogeneous parameters \\
232  & new stars with homogeneous parameters \\
312  & bright FGK stars without \\
     & \ \ homogeneous parameters (G $<$ 12) \\
    \hline
    \end{tabular}
  \end{center}
\label{tab_stats}
\end{table}

%FGK: Teff > 4000K

%%VAdibekyan: The plots should be with the new values in the xaxis...

\begin{figure*}

\begin{subfigure}{.5\linewidth}
  \includegraphics[width=\linewidth]{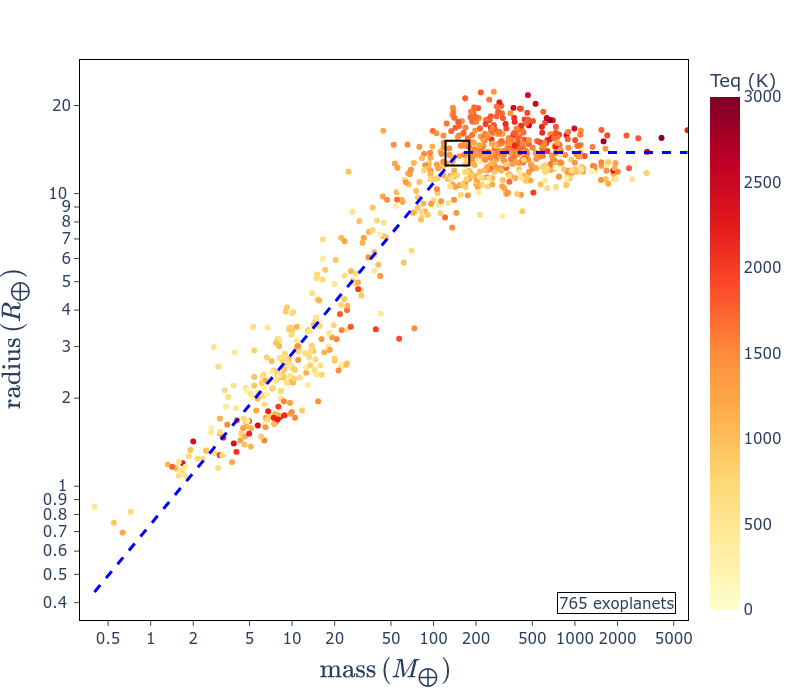}
%  \caption{All planets with hosts listed in SWEET-Cat.}
  \label{mrleft}
\end{subfigure}\hfill % <-- "\hfill"
\begin{subfigure}{.5\linewidth}
  \includegraphics[width=\linewidth]{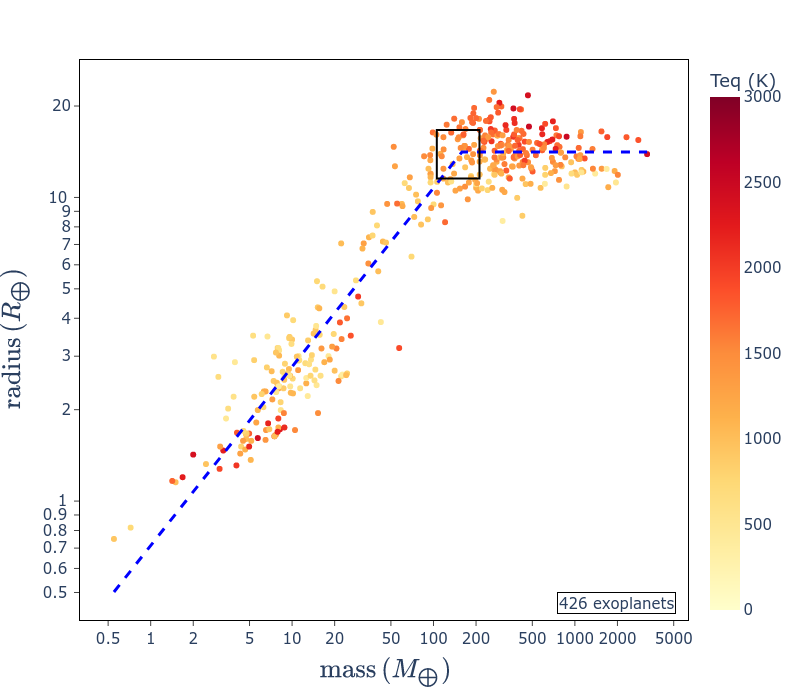}
%  \caption{Planets with hosts listed in SWEET-Cat with homogeneous spectroscopic parameters.}
  \label{mrright}
\end{subfigure}
\caption{The planetary mass - radius relation for planets orbiting stars listed in SWEET-Cat catalog. The left panel includes all the planets with precise values for planetary mass and radius, while in the right panel, are only the planets hosted by stars with homogeneous stellar parameters. The black box represents the position and its respective uncertainty for the transition between the two regimes in the diagram.}
\label{fig_mass_radius}
\end{figure*}

\subsection{SWEET-Cat planet-host statistics}

Table \ref{tab_stats} presents the updated numbers of host stars in SWEET-Cat. The percentage of stars with spectra and homogeneous parameters increases for bright stars listed in SWEET-Cat. With the work presented in this paper, despite the addition of 232 new stars with homogeneous parameters, we continue to have $\sim$86\% and $\sim$77\% completeness for FGK stars with G<9 and G<12, respectively. Completeness represents the number of host stars with homogeneous spectroscopic parameters relative to the total number of stars listed in SWEET-Cat. The continuous increase of new host stars in the catalog keeps the percentage for bright stars almost the same. In Table \ref{tab_stats} we can also see that there is still a subset of bright FGK planet-host stars (312 stars, about $\sim$25\% of the bright FGK stars) for which we did not obtain (yet) good quality data at the time of the search for this work. These will be the focus of future SWEET-Cat updates.

\section{The planetary mass-radius relation}

A brief overview of the planetary mass radius relation can be explored with the known exoplanets with both derived mass and radius. Several works have explore the mass-radius relation of exoplanets \citep[e.g.][]{Bashi-2017, Chen-2017, Ulmer-Moll-2019, Otegi-2020, Edmondson-2023}. In these works the M-R relation is characterized by fitting into different regimes that correspond to different planet populations. In some cases there are also suggestions of different correlations with other planetary system's properties which could then be explained by different theoretical arguments. It is out of the scope of this work to discuss theories that could explain the proposed correlations, but instead our only goal is to give examples and demonstrate how SWEET-Cat can be very useful to update and/or strengthen previously identified observational correlations between different parameters. In particular we want to explore the planetary mass radius relation dependence with a more homogeneous derivation of the planet equilibrium temperature which could be estimated using SWEET-Cat's homogeneous effective temperatures derived for their host stars.

%[[Variables]]  All: 765 (density < 12)
%    m1:  0.58337625 +/- 0.00913451 (1.57%) (init = 0.5)
%    b1: -0.12958799 +/- 0.01248979 (9.64%) (init = 0)
%    m2:  1.8319e-15 +/- 3.6457e-06 (199014039940.60%) (init = 0)
%    b2:  1.14029177 +/- 0.04236908 (3.72%) (init = 0)
%    xc: 2.18+/-0.08 (150+/-29)     % mass transition
%    rc: 1.14+/-0.04 (13.8+/-1.3)   % radius transition

%[[Variables]] Homogeneous: 426
%    m1:  0.58906103 +/- 0.01265397 (2.15%) (init = 0.5)
%    b1: -0.14639320 +/- 0.01711207 (11.69%) (init = 0)
%    m2:  7.8450e-11 +/- 0.02244588 (28611689857.33%) (init = 0)
%    b2:  1.14960357 +/- 0.06165980 (5.36%) (init = 0)
%    xc:  2.20+/-0.15 (158.5+/-53)  % mass transition
%    rc:  1.15+/-0.08 (14.1+/-2.6)  % radius transition

%\subsection{Equilibrium Temperature}

We start by cross-matching the updated SWEET-Cat table with the planet properties that one can find in the NASA exoplanet archive as well as the Extrasolar Planets Encyclopedia. This was easily done by following the Python tutorial available on SWEET-Cat's website\footnote{https://sweetcat.iastro.pt/catalog/SC-Python-Tutorial.pdf}. Here we collect first the planet properties listed as default in the NASA exoplanet archive \footnote{Priority was given to the NASA exoplanet archive because it provides a more complete set of planet proprieties. It also provides a column with the equilibrium temperature for each planet which we could use directly for comparison.}. If for a specific planet there are no values found in the NASA exoplanet archive, then we take the planet's properties from the Extrasolar Planets Encyclopedia. This allows to have a more complete set of planet properties.

Then the next step was to derive the planet's equilibrium temperature which is useful to compare similar planets orbiting different stars with different periods. To derive the equilibrium temperature for each planet we used Equation 3.9 from \citet[][]{Seager-2010}:
\begin{equation}
T_{\rm eq} = T_{\mathrm{eff},\ast} \left( \frac{R_{\ast}}{a}\right)^{1/2} [ \,f^{\prime}(1 - A_B)]\,^{1/4},
\end{equation}
using the effective temperature ($T_{\mathrm{eff},\ast}$) and stellar radius ($R_{\ast}$) listed in SWEET-Cat, and the semi-major axis ($a$) as the proxy for the average distance of the planet. Moreover it was assumed a bond albedo ($A_B$) of zero in the process for all planets for this estimation\footnote{Although we may expected to have different bond values for different planets, as we observe in our own solar-system planets, there are only very few measurements available for known exoplanets. Therefore since our goal will be to compare our results with others in the literature we follow the same assumption made for this and use a zero value for the albedo of all planets in our sample.}. $f^{\prime}$ is the redistribution factor with the value of $1/4$, assuming that the stellar radiation is uniformly distributed around the exoplanet.

At this point we take a sub-sample of the planets that have relative precise values for both their mass and radius. We keep only the planets with masses and radius with relative uncertainties below 45\%, and 15\% respectively. Moreover we also only consider the planets with measured velocity semi-amplitude k. The goal is to make sure that we avoid planet masses determined only by transit time variations (see e.g. \citet[][]{Adibekyan-2024}, and references therein, for a discussion). Using these constraints we get a total of 765 exoplanets. From these when considering only the planet-host with homogeneous spectroscopic parameters listed in SWEET-Cat the sample is reduced to 426 planets.

\begin{table}[t]
\caption[]{Planetary mass radius linear regimes.}
  \begin{center}

    \begin{tabular}{c|rr}
    \hline
    \hline
    Parameter     &   Full set (765)  & Homogeneous (426)  \\
    \hline
    $m_1$  &   $ 0.58  \pm  0.01$  & $ 0.59 \pm 0.01$ \\ 
    $b_1$  &   $-0.13  \pm  0.01$  & $-0.15 \pm 0.02$ \\
    $m_2$  &   $ 0.00  \pm  0.01$  & $ 0.00 \pm 0.02$ \\
    $b_2$  &   $ 1.14  \pm  0.04$  & $ 1.15 \pm 0.06$ \\
$x_c$ $[M_J]$& $ 2.18  \pm  0.08$  & $ 2.20 \pm 0.15$ \\
$x_c$ $[M_\oplus]$  &  $150  \pm  29$ &  $159  \pm  53$ \\
$y_c$ $[R_\oplus]$  &  $13.8 \pm 1.3$  &  $14.1 \pm 2.6$ \\
    \hline
    \end{tabular}
  \end{center}
\label{tab_regimes}
{\raggedright Note: $m_1$ and $b_1$ is the slope and intercept of the linear model for the low-mass planets. $m_2$ and $b_2$ is the slope and intercept of the linear model for the high-mass planets. $x_c$ and $y_c$ are the interception mass, and radius, point of the two linear models.\par}
\end{table}

Figure \ref{fig_mass_radius} presents the mass-radius relation for the two sub-samples of planets where the color of the points represents their equilibrium temperature. In both panels it is evident the presence of two clear regimes hinting at two distinct populations of planets. Both panels of Figure \ref{fig_mass_radius} show the best fitted lines minimizing a model composed of two linear regimes separated on a common mass/radius point ($x_c$). To find the best fit for this simple bilinear model, we used the \textit{lmfit} python package \citep[][]{Newville-2014}.
In the left panel the best fit shows a separation of the different populations at $x_c = 150 \pm 29 M_\oplus$ with a slope of $m_1 = 0.58 \pm 0.01$ for the low mass planets, while the slope is zero for the massive planets. Almost an equal result was obtained when fitting the points in the right panel of the figure where we find the separation between the two population of planets at $x_c = 159 \pm 53 M_{\oplus}$. The parameters obtained by this two regime model are reported in Table \ref{tab_regimes}. These values are also close to the ones reported in the work of \citet[][]{Bashi-2017} where it was used a much smaller sample (238 exoplanets) and presented slopes of $m_1 = 0.55$ for the small planets, $m_2 = 0.01$ for the large planets, and a transition point of $x_c = 124 M_\oplus$. 
%where it presents slopes of $m_1 = 0.55 \pm 0.02$ for the small planets, $m_2 = 0.01 \pm 0.02$ for the large planets, and a transition point of $x_c = 124 \pm 7 M_\oplus$. 

\begin{figure*}

\begin{subfigure}{.5\linewidth}
  \includegraphics[width=\linewidth]{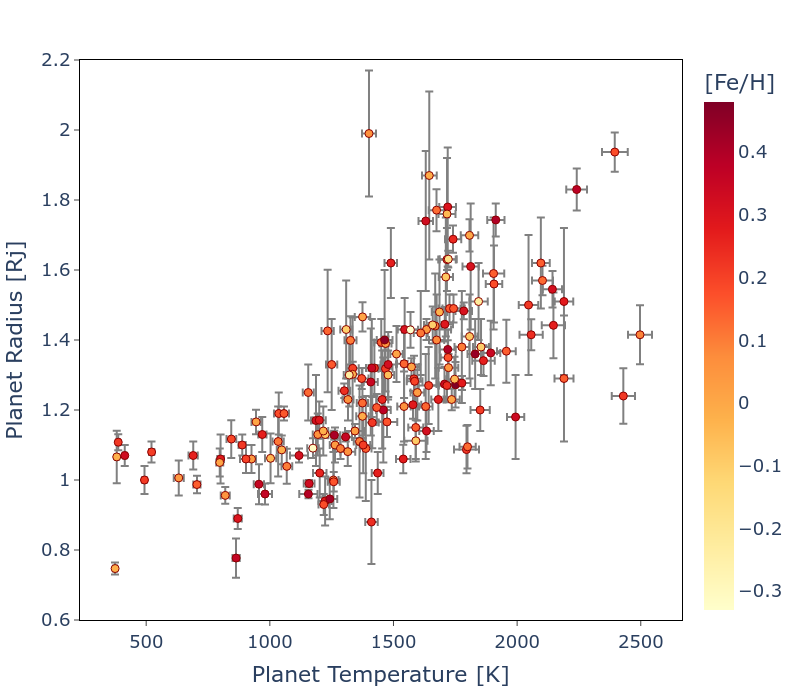}
  %\caption{}
%  \label{RadTeff}
\end{subfigure}\hfill % <-- "\hfill"
\begin{subfigure}{.5\linewidth}
  \includegraphics[width=\linewidth]{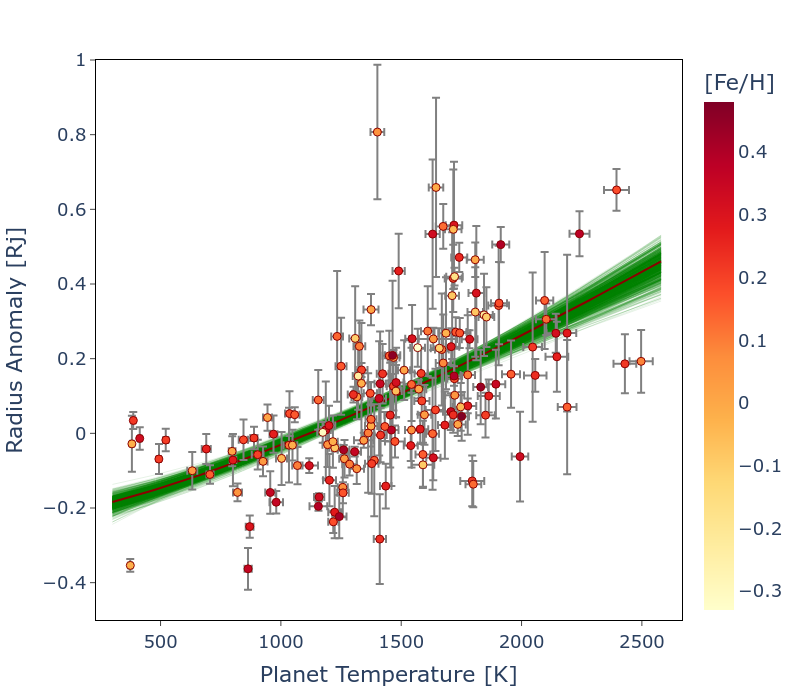}
  %\caption{Planet radii vs. equilibrium temperature for the 212}
%  \label{RadATeff}
\end{subfigure}

\caption{Left: The planet radius and equilibrium temperature for the 159 massive exoplanets ($0.5 M_J <$ mass $< 10 M_J$) with homogeneous high precision [Fe/H]. Right panel is the radius anomaly vs. the equilibrium temperature for the same planets. The line is the best fit of a power law.}
\label{fig_radTeq}
\end{figure*}

Other works also presented consistent slopes and transition mass between these two populations, although some of these also identify an additional transition at lower masses at around $x_{c2} = 4.37 \pm 0.72 M_\oplus$ \citep[e.g.][]{Muller-2023, Otegi-2020, Chen-2017}. This additional transition of regimes is not clearly seen in our Figure \ref{fig_mass_radius}, although there is a hint for this at the lower mass regime. Looking carefully one can notice the increase of hotter planets below $\sim 10 M_\oplus$ and below $\sim 2 R_\oplus$ which even seems to follow a different slope when compared with the rest of the surrounding planets in the figure. This might be connected with the known radius gap that is discussed in many different works \citep[e.g.][]{Fulton-2017, Kubyshkina-2022}. We note however that several planets that would fall at these low-mass regime, such as the ones from the Trappist system \citep[e.g.][]{Gillon-2017a}, were excluded because the semi-amplitude velocity k was not reported in the databases. This comes from our sample selection to try to avoid masses that were not derived by the radial-velocity method\footnote{Although this is correctly assumed for most of the planets, there are a few stars listed in the exoplanets databases for which no radial velocity amplitude is reported even though there are masses values in literature derived by the RV method (e.g. WASP-43)}. In addition most of these excluded planets orbit faint host stars and are therefore more difficult to have spectroscopic data and thus unlikely to have homogeneous parameters in SWEET-Cat. Because of these reasons, the third population of small planets should be dealt with caution given the general lower precision of its stellar and planetary properties.

A clearly visible color gradient can be observed for giant planets in Figure \ref{fig_mass_radius} indicating a much stronger positive correlation with the equilibrium temperature rather than with mass. This was already noticed in previous works, for example in \citet[][]{Edmondson-2023} where the planet radius was fitted by a power law both on planet mass and the equilibrium temperature (equation 9 of their paper: $\frac{R}{R_{\oplus}} = C T_{\rm eq}^{\beta_1} \left( \frac{M}{M_{\oplus}} \right)^{\beta_2}$, where $C$, $\beta_1$ and $\beta_2$ are constants and $T_{\rm eq}$ is the equilibrium temperature of the planet). \citet[][]{Edmondson-2023} derived a temperature index $\beta_1 = 0.35 \pm 0.02$  and a mass index of $\beta_2 = 0.00 \pm 0.01$. Doing a similar fit to our data in Figure \ref{fig_mass_radius} for planets with masses above 150 $M_\oplus$, in the left panel, we get a temperature index $\beta_1 = 0.31 \pm 0.01$  and a mass index of $\beta_2 = -0.06 \pm 0.01$. Doing the same fit but for the planets with homogeneous spectroscopic stellar parameters, with masses above 159 $M_\oplus$, we get a temperature index $\beta_1 = 0.40 \pm 0.01$  and a mass index of $\beta_2 = 0.00 \pm 0.01$. This represents a very substantial difference for the temperature index when comparing both samples. We note that the sampling distributions resulting from this combined fit are close to Gaussian distributions and are quite narrow as demonstrated by the adopted parameter errors. We also note a small correlation noticeable between the parameters which is more evident between $C$ and $\beta_1$. The conclusion here is that the correlation with the temperature is stronger when using the smaller sample of planets with the more precise homogeneous spectroscopic parameters which is relevant for the computation of the equilibrium temperature of the planet. %(add something more???)
%[Add some physical intrepretation here? "Inflation".]
%Thought: Isn't expected that hotter planets, will have higher densities, which in turn can contrabalance more the gravity?

% All, at 150
%[-0.06252289  0.30979972  0.33426826]
%[0.00347401 0.00413998 0.01537523]

%Homogeneous at 159
%[-1.23391127e-04  3.94935553e-01 -9.50180348e-02]
%[0.00535922 0.00700137 0.02933276]

%\subsection{Revisiting the radius anomaly of Jupiter-like planets}
\section{Revisiting the radius anomaly of Jupiter-like planets}

%\begin{figure*}

%\begin{subfigure}{.475\linewidth}
%  \includegraphics[width=\linewidth]{radiusAnomaly_corrected_metalicity_massive.png}
  %\caption{}
%  \label{RadTeff}
%\end{subfigure}\hfill % <-- "\hfill"
%\begin{subfigure}{.475\linewidth}
%  \includegraphics[width=\linewidth]{radiusAnomaly_corrected_metalicity_massive_nasa.png}
  %\caption{Planet radii vs. equilibrium temperature for the 212}
%  \label{RadATeff}
%\end{subfigure}

%\caption{Radius anomaly with equilibrium temperature trend corrected versus the metallicity of the host star [Fe/H]. Left: Using the homogeneous parameters. Right: Using parameters listed in the NASA exoplanet archive. }
%\label{fig_anomaly_metallicity}
%\end{figure*}

The radius dependence on equilibrium temperature for massive planets was previously presented in the work of \citet[][]{Laughlin-2011} where they investigated the radius anomaly observed in giant exoplanets. These planets, despite being expected to follow a mass-radius relationship consistent with a core made of heavy elements and a hydrogen-helium envelope, exhibit anomalously large radii. Potential mechanisms for these inflated radii, including tidal heating, enhanced atmospheric opacities, and heat redistribution were explored. Understanding which mechanism dominates is crucial for explaining the observed deviations. One of the interesting results presented in the work of \citet[][]{Laughlin-2011} was the evidence for a significant correlation between the residual radius anomalies and host star metallicities. A correlation that is generally interpreted as evidence that metal-rich protoplanetary disks lead to planets with larger core masses and thus with larger radii.

%Check the upper mass limit for the selection of massive planets.
In this section we will quickly revisit the relations presented in \citet[][]{Laughlin-2011} using the updated SWEET-Cat data with the only goal to verify the metallicity correlation on the residual radius anomalies. While in \citet[][]{Laughlin-2011} the planets were selected to have masses between 0.1 and 10 Jupiter masses here we use our mass transition in the previous section by selecting planets with masses above 159 earth masses (0.5 Jupiter masses). From these we only consider the planets hosted by stars with homogeneous stellar parameters and with a good precision below or equal to 0.05 dex in [Fe/H]. This leaves us with only 158 planets, but with very precise and homogeneous metallicity determinations. The left panel of Figure \ref{fig_radTeq} shows the dependence of the radius on equilibrium temperature for these massive planets. We also computed the radius anomaly following the same procedure as in \citet[][]{Laughlin-2011} by using its equation 3 that provides an acceptable approximation to the \citet[][]{Bodenheimer-2003} baseline structural models throughout the mass and temperature ranges in this sample of massive planets:
\begin{equation}
\begin{aligned}
R_{\rm pl}/R_{\rm Jup} &= 1.08417+0.0940857\,m - 0.242831\,m^{2} \\
                       &+0.0947349\,m^{3} +0.0387851\,t+0.00243981\,mt \\
                       &-0.0244656\,m^{2}t +0.0130659\,m^{3}t +0.0240409\,t^{2} \\
                       &-0.0419296\,mt^{2} +0.00693348\,m^{2}t^{2} \\
                       &+0.00302157\,m^{3}t^{2} \, ,
\end{aligned}
\end{equation}
The radius anomaly is computed by subtracting the result of this fit to the observed radius values for all the planets. The right panel of Figure \ref{fig_radTeq} reproduces Figure 2 of \citet[][]{Laughlin-2011} for our sample of massive planets around precisely characterized host stars in SWEET-Cat. In this panel we also used a power law to fit the distribution points (${\cal R}\propto T_{\rm eq}^{\alpha}$, ${\cal R}$ being the radius anomaly, and $T_{\rm eq}$ the equilibrium temperature)\footnote{We note that a linear fit to these points will do basically the same work, nevertheless we keep the fit of the same power law model for a direct comparison with the original work.} for which we derive an $\alpha=1.35 \pm 0.51$ which is consistent with the value of $\alpha=1.4\pm0.6$ originally presented in their work with only 90 early detected transiting massive exoplanets. 

%samples_marg.shape  Derived linear fit considering errors on x and y
%print(np.mean(samples_marg, axis=0))
%print(np.std(samples_marg, axis=0))
%[-0.41542548  0.04010424]
%[0.02480114 0.00610872]

\begin{figure}
  \centering
  \includegraphics[width=\linewidth]{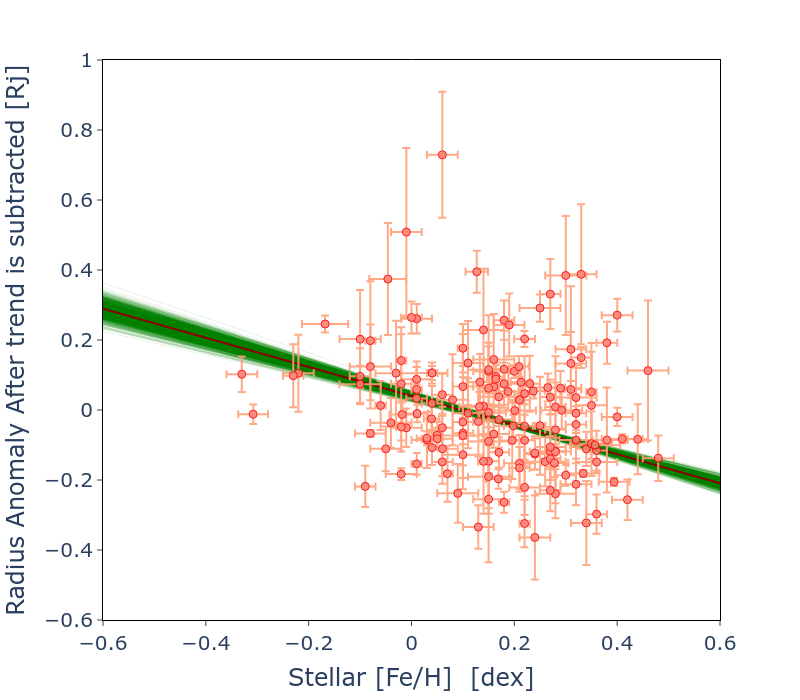}
  \caption{Radius anomaly with equilibrium temperature trend corrected versus the metallicity of the host star [Fe/H].}
  \label{fig_anomaly_metallicity}
  \includegraphics[width=\linewidth]{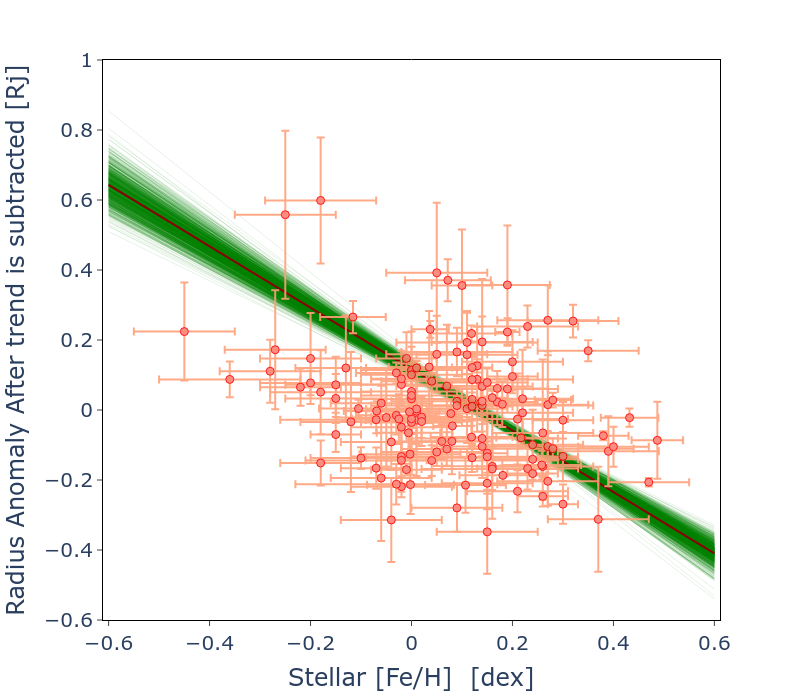}
  \caption{Same as Figure \ref{fig_anomaly_metallicity} but using parameters directly listed in the NASA exoplanet archive.}
  \label{fig_anomaly_metallicity_nasa}
\end{figure}

%samples_marg.shape  Derived linear fit considering errors on x and y
% parameters coming from literature NASA exoplanet archive
%print(np.mean(samples_marg, axis=0))
%print(np.std(samples_marg, axis=0))
%[-0.87747728  0.11668378]
%[0.05627708 0.01089849]
After correcting for the equilibrium temperature trend and plotting the resulting radius anomaly against the host stellar metallicity, \citet[][]{Laughlin-2011} reported an evidence for a negative correlation in its Figure 3, but no slope value was presented. Here we reproduce this correlation using our sample of planets with precise and homogeneous metallicity in Figure \ref{fig_anomaly_metallicity} for which we find a negative slope of $ m = -0.42 \pm 0.04 $ (we also derived a Spearman correlation coefficient of -0.23 with a low p-value of 0.004 for these data points) which confirms the observed trend originally presented.

%We want to note that this exercise was done also considering different and larger samples of massive planets in order to check if this would result in significant differences. Namely we allowed planets with less precision on the host's metallicity (0.10 dex) and also considered the inclusion of planets whose hosts do not have homogeneous metallicities in SWEET-Cat (increased the sample to a total of 328 planets). For this larger sample of planets the fitted power law for the radius anomaly provides a $\alpha=1.39 \pm 0.37$ and the correlation with the metallicity is very similar with a value of $m = -0.24 \pm 0.04 $ with a much smaller p-value of $3.8e^{-5}$ presenting a metallicity dependence fully consistent with the values reported for the smaller and more precise sample of massive planets.

As a comparison to the use of our homogeneous SWEET-Cat parameters, we did the same exercise using the parameters that are found directly in the NASA exoplanet archive for the same 158 planets. For these we collect directly the equilibrium temperature and the stellar metallicity compiled from literature. For 14 of these exoplanets the stellar metallicity was not available in the NASA database. For the few other exoplanets that did not had uncertainties for the stellar metallicity we assumed an error of 0.10 dex. For the few exoplanets with no equilibrium temperature reported in the NASA exoplanet database we used equation 1 with the effective temperature and stellar radius reported in the NASA's database to estimate the missing values. With these data at hand the power law fit on the radius anomaly vs. equilibrium temperature gave consistent results with a very similar $\alpha=1.33 \pm 0.51$ meaning that the changes in equilibrium temperature does not affect much these results. However when we observe the metallicity dependence on the radius anomaly with the equilibrium temperature trend corrected (Figure \ref{fig_anomaly_metallicity_nasa}) we derived a significant different linear fit, with a stronger negative slope $ m = -0.88 \pm 0.12 $ which is also less precise (we also derived a Spearman correlation coefficient of -0.21 with a p-value of 0.01 for these data points). Moreover, we derived Spearman correlations coefficient distributions to take into account the errors included for both Figures 4 and 5. The average correlation coefficient obtained was -0.20 with an average p-value of 0.03 for Figure 4, while for Figure 5 we obtained a similar correlation of -0.18 but with a significant higher average p-value of 0.07. The higher uncertainty of the fit as well as the higher values of p-value obtained for Figure 5 are caused by the higher uncertainties of the metallicities available in the literature.

%no errors
%    k:     1.0150e-05 +/- 4.0694e-05 (400.94%) (init = 1)
%    m:     1.41613422 +/- 0.50373967 (35.57%) (init = 1)
%    mfeh: -0.23520846 +/- 0.08640409 (36.74%) (init = 1)
%    bt:   -0.17574918 +/- 0.10312511 (58.68%) (init = 1)

%errors with monte carlo
%mfeh: -0.22665702156824455 0.047902762279017276
%m: 1.401230661501942 0.17270260410257457
%k: 2.756240389341663e-05 5.462348287186577e-05
%bt: -0.18120373747658836 0.029800541958665525

We also performed a simultaneous 3D fit to the radius anomaly with dependence on both the equilibrium temperature and metallicity. The goal was to simply check if the observed correlations remained consistent with the step-by-step analysis done as in \citet[][]{Laughlin-2011}. Therefore we used a model to follow the relation ${\cal R} = k T_{\rm eq}^{\alpha} + m {\rm[Fe/H]} + b$ for which we derived the following coefficients: $k = 2.75\mathrm{e}^{-5} \pm 5.46\mathrm{e}^{-5}$, $\alpha = 1.40 \pm 0.17$, $m = -0.23 \pm 0.05$, and $b = -0.18 \pm 0.03$. We note that there is a significant decrease in the uncertainty of $\alpha$ doing this simultaneous fit. This way we are able to confirm that both trends on equilibrium temperature and metallicity remain present, although the metallicity slope seems to have reduced significantly.

All these exercises done on updated exoplanet data confirm that the metallicity trend is present. Using homogeneous data we are able to obtain a more precise correlation, although depending on the approach used to extract this dependence provide different slope values. This metallicity dependence may be connected with the presence of larger and more massive cores for the planets orbiting metal-rich stars, which could compensate the inflation of their radius. We recall that the baseline models used in \citet[][]{Laughlin-2011} come from \citet[][]{Bodenheimer-2003} which assumed a near-solar composition. It would be interesting to compare with more updated models, preferably with the metallicity accounted for, but this is out of the scope of this paper.

\section{Summary}

We present another large update of SWEET-Cat where we derive precise and homogeneous spectroscopic stellar parameters for 232 new planet-host stars. To achieve this we compiled spectra collected from different instruments, by both exploring public archives and making using of our own observation programs at different observatories. The spectroscopic parameters were complemented with additional data as in previous works, including estimation of stellar masses, radius, and trigonometric surface gravities which are made available at the SWEET-Cat website. 

For all new planet-hosts added in SWEET-Cat we maintain the cross reference columns for an easy correlation with the planets's properties listed in the NASA archive exoplanet database as well as the Extrasolar Planets Encyclopedia.

We make a quick review of the planetary mass radius relation of exoplanets using homogeneous spectroscopic data where we can easily identify the two populations of exoplanets following two different linear relations for low-mass and high-mass planets. The relations derived in this work are consistent with several others derived in different works, including the dependence on equilibrium temperature observed for the massive planets.

We also revisit the radius anomaly observed for giant planets with the goal to quantify the metallicity correlation presented as evidence in \citet[][]{Laughlin-2011}. Using different larger samples of massive exoplanets, we find that the metallicity correlation persists with the same models. However, these studies assume that giant planets are perfectly absorbing (black) bodies, which is unrealistic. Early exoplanet models have demonstrated that the Bond albedo depends on the spectral type of the host stars and the expected composition of planetary atmospheres, considering their temperatures \citep[e.g.][]{Gelino-1999, Sudarsky-2000}. Recent research has shown that photochemical hazes, rather than cloud condensates, may serve as the primary aerosols depending on the planets' temperatures \citep[][]{Steinrueck-2023}, thereby altering their expected albedos. Consequently, more comprehensive and recent models should be examined to assess the significance of this metallicity correlation on the radii of these massive planets.

We will continue to expand SWEET-Cat in the following years to provide the community with a more complete data set relevant for statistical studies of exoplanets. There is still a significant number ($\gtrsim$ 300) of relatively bright planet-host listed without homogeneous parameters. We will provide spectroscopic parameters once good quality spectra are available. Furthermore, we have plans to include additional stellar parameters in SWEET-Cat. These include $v\sin i$, stellar activity indices, homogeneous element abundances for the stars with the highest quality spectra, make the spectra available for download on the website, and include more robust determinations of stellar masses, radius, and ages from stellar modeling.

\begin{acknowledgements}

This  work  was  supported  by FCT - Fundac\c{c}\~ao  para  a  Ci\^encia  e  Tecnologia through national  funds and by FEDER through COMPETE2020 - Programa Operacional Competitividade e Internacionaliza\c{c}\~ao by these grants: 2022.06962.PTDC; UIDB/04434/2020; UIDP/04434/2020. S.G.S acknowledges the support from FCT through Investigador FCT contract nr. CEECIND/00826/2018 and  POPH/FSE (EC). E.D.M. further acknowledges the support from FCT through the Stimulus FCT contract 2021.01294.CEECIND. Co-funded by the European Union (ERC, FIERCE, 101052347). BMTBS is supported by an FCT fellowship, grant number 2022.11805.BD. Views and opinions expressed are however those of the author(s) only and do not necessarily reflect those of the European Union or the European Research Council. Neither the European Union nor the granting authority can be held responsible for them. 
This research has made use of the NASA Exoplanet Archive, which is operated by the California Institute of Technology, under contract with the National Aeronautics and Space Administration under the Exoplanet Exploration Program. This work has made use of data from the European Space Agency (ESA) mission {\it Gaia} (\url{https://www.cosmos.esa.int/gaia}), processed by the {\it Gaia} Data Processing and Analysis Consortium (DPAC, \url{https://www.cosmos.esa.int/web/gaia/dpac/consortium}). Funding for the DPAC has been provided by national institutions, in particular the institutions participating in the {\it Gaia} Multilateral Agreement. This work has been carried out within the framework of the NCCR PlanetS supported by the Swiss National Science Foundation. SH acknowledges CNES funding through the grant 837319. AM acknowledges funding from a UKRI Future Leader Fellowship, grant number MR/X033244/1. B.R-A acknowledges funding support from ANID Basal project FB210003.

\end{acknowledgements}

\bibliographystyle{aa}
\bibliography{sousa_bibliography}

\end{document}